\documentclass[aps,prl,twocolumn,showpacs,superscriptaddress]{revtex4}

\usepackage{graphicx} % Include figure files
\usepackage{longtable}

\begin{document}

\title
{Half-metallic properties of atomic chains of carbon-transition
metal compounds}

\author{S. Dag}
\affiliation{Department of Physics, Bilkent University, Ankara
06800, Turkey}
\author{S. Tongay}
\affiliation{Department of Physics, Bilkent
University, Ankara 06800, Turkey}
\author{T. Yildirim}
\affiliation{NIST Center for Neutron Research, National Institute
of Standards and Technology, Gaithersburg MD 20899}
\author{E. Durgun}
\affiliation{Department of Physics, Bilkent University, Ankara
06800, Turkey}
\author{R. T. Senger}
\affiliation{Department of Physics, Bilkent University, Ankara
06800, Turkey}
\author{C. Y. Fong}
\affiliation{Department of Physics, University of California,
Davis, CA 95616, USA}
\author{S. Ciraci}\email{ciraci@fen.bilkent.edu.tr}
\affiliation{Department of Physics, Bilkent University, Ankara
06800, Turkey}

\date{\today}

\begin{abstract}
We found that magnetic ground state of one-dimensional atomic
chains of carbon-transition metal compounds exhibit half-metallic
properties. They are semiconductors for one spin-direction, but
show metallic properties for the opposite direction. The spins are
fully polarized at the Fermi level and net magnetic moment per
unit cell is an integer multiple of Bohr magneton. The
spin-dependent electronic structure can be engineered by changing
the number of carbon and type of transition metal atoms. These
chains, which are stable even at high temperature and some of
which keep their spin-dependent electronic properties even under
moderate axial strain, hold the promise of potential applications
in nanospintronics.

\end{abstract}


%72.25.-b Spin polarized transport
%71.70.Ej Spin-orbit coupling, Zeeman and Stark splitting, Jahn-Teller effect
%73.40.Jn Metal-to-metal contacts
%75.50.Cc Other ferromagnetic metals and alloys
%71.20.Be Transition metals and alloys

%73.22.-f   Electronic structure of nanoscale materials:  clusters,
%            nanoparticles, nanotubes, and nanocrystals
%61.48.+c   Fullerenes and fullerene-related materials
%73.20.Hb   Impurity and defect levels; energy states of adsorbed species
%71.30.+h   Metal-insulator transitions and other electronic transitions
%71.20.Tx   Fullerenes and related materials; (Band str.)
%71.15.Nc   Total energy and cohesive energy calculations

\maketitle

Spin-dependent electronic transport has promised  revolutionary applications using
giant-magneto-resistance in magnetic recording and nonvolatile memories
\cite{prinz,ball,wolf}. Half-metals (HM) \cite{groot,pickett} are a class of materials, which
exhibit spin-dependent electronic properties relevant to spintronics. In HMs,  due to broken
spin-degeneracy, energy bands $E_{n}({\bf k },\uparrow)$ and $E_{n}({\bf k},\downarrow)$
split  and each band accommodates one electron per \textbf{k}-point. Furthermore, they are
semiconductor for one spin direction, but show metallic properties for the opposite spin
direction. Accordingly, the difference between the number of electrons of different
spin-orientations in the unit cell, $N=N_{\uparrow}-N_{\downarrow}$, must be an integer and
hence the spin-polarization at the Fermi level,  $P=D(E_{F},\uparrow)-D(E_{F},\downarrow)/
[D(E_{F},\uparrow)+D(E_{F},\downarrow)]$,  is  complete \cite{pickett}. This situation is in
contrast with the ferromagnetic metals, where both spin-directions contribute to the density
of states at $E_{F}$ and spin-polarization $P$ becomes less than 100\%. Even though
three-dimensional (3D) ferromagnetic Heusler alloys and transition-metal oxides exhibit HM
properties \cite{park}, they are not yet appropriate for spintronics because of difficulties
in controlling stoichiometry  and the defect levels destroying the coherent spin-transport.
Zinc-blende (ZB) HMs with high magnetic moment $\mu$ and high Curie temperature $T_{c}>400 K$
(such as CrAs, and CrSb in ZB structure) have been grown only in thin-film forms
\cite{akinaga}. More recently,  it has been predicted that four new ZB crystals can be HM at
or near their respective equilibrium lattice constants \cite{pask}.

In this letter, we report that very simple and stable one-dimensional (1D) structures, such
as linear atomic chains of carbon-transition metal compounds, \textit{i.e.} C$_{n}$(TM), show
half-metallic properties. The prediction of half-metallic behavior in 1D atomic chains is new
and of fundamental interest, in particular in the field of fermionic excitations with spin
degree of freedom. Besides, the present finding may lead to potential applications in the
rapidly developing field of nanospintronics, such as tunnelling magnetoresistance, spin
valve, and nonvolatile magnetic devices.

In  earlier transport studies, the spin-direction of conduction electrons was generally
disregarded, in spite of the fact that spin orientation of electrons decays much slower than
their momentum \cite{wolf}. Magnetic ground state of TM-adsorbed single-wall carbon nanotubes
(SWNT) \cite{dag,yang}, spontaneous spin-polarized electron transport through finite TM wires
\cite{rodrigues}, and oscillatory spin-polarized conductance and spin-valve effect through
finite carbon wires capped with Co atoms in  between gold electrodes \cite{pati} have been
treated recently. However, half-metallicity predicted in periodic C$_{n}$(TM) is a behavior
fundamentally different from those magnetic properties found in earlier systems in Ref.
~\cite{dag,yang,rodrigues,pati} and is a novel feature of 1D systems that allows both
semiconducting and metallic properties coexisting in the same structure.

Our predictions are obtained from the first-principles pseudopotential plane wave
calculations within DFT using generalized gradient approximation (GGA) and ultrasoft
pseudopotentials \cite{vasp}. C$_{n}$(TM) chains have been treated in  supercell geometry,
and Brillouin zone is sampled by $10-80$ special \textbf{k}-points depending on the supercell
size. All the atomic positions, and the supercell lattice parameter $c$ along chain axis are
optimized by minimizing the total energy, $E_{T}^{sp}$, the forces on the atoms, as well as
the stress of the system calculated by spin-polarized (or spin-relaxed) GGA. Since $\Delta
E=E_{T}^{su}- E_{T}^{sp}$ (where $E_{T}^{su}$ is spin-unpolarized total energy) and net
magnetic moment $\mu$ are both positive, these compound chains have ferromagnetic ground
state ~\cite{spinrel}. While our study has covered a large family of C$_{n}$(TM) chains, our
discussion will focus on C$_{n}$Cr.

\begin{figure}
\includegraphics[scale=0.42]{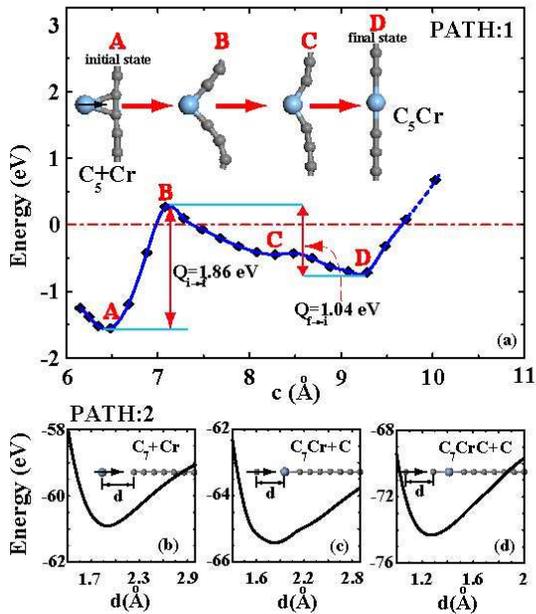}
\caption{ Transition state analysis for two different reaction paths (a) Path-1: Variation of
energy, $E_{T}(c)$, for the transition from the C$_{5}$+Cr initial state to the linear
C$_{5}$Cr HM as final state: Q$_{i\rightarrow f}$ and Q$_{f\rightarrow i}$ are energy
barriers involved in the transitions. Zero of energy is taken relative to the free Cr atom
and free periodic C-linear chain.  (b)-(d) Path-2: Variation of interaction energy with
distance $d$ between C$_{7}$ and Cr, C$_{7}$Cr and C, C$_{7}$CrC and C for adatom attaching
from one end. } \label{fig:trans}
\end{figure}

Finite size linear chain of single C atom has already been synthesized experimentally
\cite{roth}. The double bond between carbon atoms and doubly degenerate $\pi$-band crossing
the Fermi level underlie the stability of the chain and its unusual electronic properties. We
examine the formation of linear C$_{5}$Cr HM by performing  transition state analysis along
two different reaction paths. Normally, a Cr atom is attracted by C-linear chain and
eventually forms a bridge bond over a C-C bond. We take this bound state (specified as
C$_{5}$+Cr) as an initial state of the first reaction path for transition to the final state
corresponding to the linear C$_{5}$Cr HM as illustrated in
Fig.~\ref{fig:trans}(a)\cite{energy}. The energy barrier necessary to go from the initial
state to the final state is Q$_{i\rightarrow f}$=1.86 eV. Once the final state has formed, it
is prevented from going back to the initial state by a significant barrier of
Q$_{f\rightarrow i}$=1.04 eV. However, the energy barrier Q$_{i\rightarrow f}$ disappears
totally and hence the process becomes exothermic for the second reaction path where HM is
grown from one end of the chain by attaching first Cr atom then C atoms sequentially. Each
ad-atom (Cr or C) is attracted to chain and eventually becomes bound to it as shown in
Fig.~\ref{fig:trans}(b-d). This analysis lets us believe that half-metallic C$_{n}$(TM)
chains are not only of fundamental interest, but also can be realized experimentally.

Whether a periodic C$_{n}$Cr linear chain is stable or it can transforms to other structures
has been examined by an extensive investigation of Born-Oppenheimer surface. Local minima of
the total energy have been searched by optimizing the structure starting from transversally
displaced chain atoms for varying lattice parameters.  The linear chain structure has been
found to be stable and energetically favorable relative to zigzag structures. The phonon
calculations of C$_{n}$Cr, yielding positive phonon frequencies ($\Omega_{TO}(k=0)=$89, 92,
411 cm$^{-1}$: $\Omega_{LO}(k=0)=$421, 1272, 1680 cm$^{-1}$ for $n=$3 and
$\Omega_{TO}(k=0)=$13, 71, 353, 492 cm$^{-1}$; $\Omega_{LO}(k=0)=$489, 1074, 1944, 2102
cm$^{-1}$ for $n=$4) corroborate the above analysis of stability. However, for $n=$9 some of
the frequencies get negative indicating an instability for a large $n$. In addition, we
performed high temperature (T=750-1000 K) \textit{ab-initio} molecular dynamics calculations
using Nos{\'e} thermostat,  where atoms are displaced in random directions. All these tests
have provided strong evidence that the linear chain structures with small $n$ are stable. To
weaken the constraints to be imposed by supercell geometry, calculations have been done by
using double supercells including two primitive unit cells of the chains. Peierls instability
that may cause the splitting of metallic bands at the Fermi level did not occur in C$_{n}$Cr
linear chain structures. Table I summarizes the calculated magnetic and electronic properties
of C$_{n}$Cr linear chain.  We also examined how an axial strain may affect the half-metallic
behavior of these chains. HM character of C$_{4}$Cr was robust under $\epsilon=\pm 0.05$.
Small band gap C$_{5}$Cr remains HM for $\epsilon=0.05$, but is rendered a ferromagnetic
metal under $\epsilon=-0.05$. While C$_{3}$Cr changes to a semiconductor under
$\epsilon=0.10$, it becomes a ferromagnetic metal with $\mu=3.1$ under $\epsilon=-0.10$.

\begin{table}
\caption{\label{tab:ref} Results of spin-polarized
first-principles calculations for C$_{n}$Cr linear chains. $\Delta
E_T$ is the difference between spin-paired (non-magnetic) and
spin-polarized (magnetic) total energies.  $c$ is the optimized 1D
lattice parameter. $\mu$ is the total magnetic moment per unit
cell in units of Bohr magneton, $\mu_{B}$. M, S, and HM stand for
metal, semiconductor, and half-metal, respectively. By convention
majority and minority spins are denoted by $\uparrow$ and
$\downarrow$. The numerals in the last column are the band-gap
energies in eV.}
\begin{ruledtabular}
\begin{tabular}{cc|c|c|c|c|cc}
  &1D-compound & $\Delta E_{T}$(eV)& $c$(\AA) & $\mu(\mu_{B})$ & Type: $\uparrow$(eV) $\downarrow$(eV)   \\
  \hline
  &CCr& 1.8&  3.7 & 2.0 & S:$\uparrow$=0.7 $\downarrow$=1.0  \\
  \hline
  &C$_{2}$Cr& 2.8& 5.2  & 4.0 & HM: $\uparrow$=M $\downarrow$=3.3 \\
  &C$_{3}$Cr& 3.0& 6.5 & 4.0 & HM:$\uparrow$=0.4  $\downarrow$=M \\
  &C$_{4}$Cr& 3.0& 7.9 & 4.0 & HM:$\uparrow$=M  $\downarrow$=2.9 \\
  &C$_{5}$Cr& 2.5& 9.0 & 4.0 & HM:$\uparrow$= 0.6  $\downarrow$=M \\
  &C$_{6}$Cr& 3.1& 10.3 & 4.0 & HM:$\uparrow$=M  $\downarrow$=2.4 \\
  &C$_{7}$Cr& 2.5 & 11.6 & 4.0 & HM:$\uparrow$=0.5 $\downarrow$=M\\
%\hline
%  &SiCr& 2.5& 5.0 & 6.0 & HM:$\uparrow$=M $\downarrow$=0.2\\
%  &Si$_{2}$Cr&3.5& 6.7 & 4.0 & HM:$\uparrow$=M $\downarrow$=1.8\\
%  &Si$_{3}$Cr&2.8& 9.0 & 4.0 & HM:$\uparrow$=0.7 $\downarrow$=M\\
%  &Si$_{4}$Cr$_{1}$&??& 10.7 & 4.0 & HM:$\uparrow$=M $\downarrow$=??\\
%  &Si$_{5}$Cr&3.1& 13.2 & 4.0 & HM:$\uparrow$=0.6 $\downarrow$=M\\
%  &Si$_{6}$Cr& 3.4 & 15.5  & 4.0  & HM:$\uparrow$=M $\downarrow$=1.1 \\
%  &Si$_{7}$Cr&2.6& 16.8 & 4.0 & HM:$\uparrow$=0.5 $\downarrow$=M\\
\end{tabular}
\end{ruledtabular}
\end{table}

Spin-polarized electronic band structures are strongly dependent on $n$. For example, all
C$_{n}$Cr we studied are HM except CCr, which is a semiconductor. For even $n$, majority spin
bands are metallic, but minority spin bands are semiconducting with large band gaps
($E_{g,\downarrow}\sim$3 eV). This situation, however, is reversed for odd $n$, where
majority spin bands become semiconducting with relatively smaller gaps
($E_{g,\uparrow}\sim$0.5 eV), but minority bands are metallic. This even-odd $n$ disparity is
closely related to bonding patterns in different chains. For example, for odd $n=3$,
respective bond lengths are in \AA~~ -C-1.28-C-1.28-C-1.95-Cr-, and for even $n=4$,
-C-1.25-C-1.33-C-1.25-C-2.1-Cr-. It appears that, while double bonds are forming between all
atoms for odd $n$, for even $n$ triple and single bonds form alternatingly between C atoms,
and single bonds occur between C and Cr atoms with relatively longer bond lengths
\cite{pati}. Consequently, the overlap between Cr and C orbitals and hence relative energy
positions of bands vary depending on whether $n$ is even or odd. The C$_{n}$Co linear chains
exhibit also similar spin-dependent electronic properties. While CCo chain is a ferromagnetic
metal with $\mu=$0.5 $\mu_{B}$, C$_{n}$Co chains (for 2$< n <$6 studied in this work) are HM.
They have even-odd $n$ disparity; $E_{g}$ is relatively larger for even $n$ but decreases as
$n$ increases. Interestingly, the odd $n$ C$_{n}$Co chains, where nearest carbon atoms have
spins smaller but opposite to that on Co, exhibit a ferrimagnetic behavior.

\begin{figure}
\includegraphics[scale=0.45]{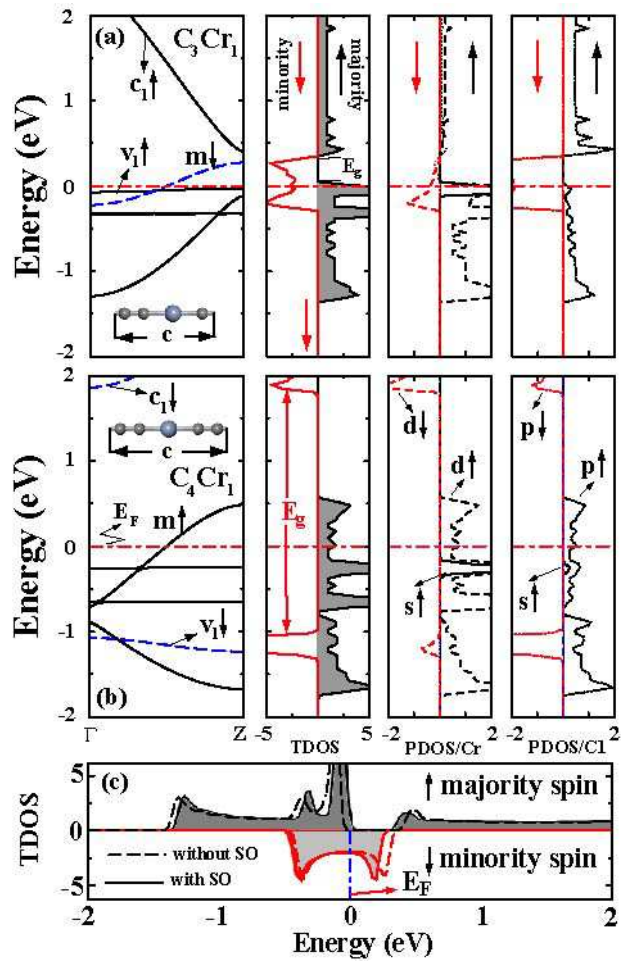}
\caption{(a) Energy band structure of C$_{3}$Cr; corresponding
total density of states (TDOS) for majority ($\uparrow$) and
minority ($\downarrow$) spins; orbital projected local density of
states at Cr atom (PDOS/Cr), at C atoms first nearest neighbor to
Cr (PDOS/C1). (b) C$_4$Cr. State densities with $s,p,d$ orbital
symmetry in PDOSs are shown by thin continuous, dotted, broken
lines, respectively. Zero of energy is set at $E_{F}$. Metallic
band crossing the Fermi level, highest valence and lowest
conduction bands are labelled by $m$, $v_{1}$ and $c_{1}$,
respectively. (c) Majority and minority TDOS of C$_{3}$Cr
calculated with and without spin-orbit coupling (SO).}
\label{fig:band1}
\end{figure}

Half-metallic electronic structure and resulting spin-dependent properties of C$_n$Cr linear
chains are shown by the bands and density of states  in Fig.~\ref{fig:band1}. The odd-even
$n$ disparity is clearly seen. The double degenerate $\pi$-band (denoted by $m\downarrow$ for
$n=$3 or $m\uparrow$ for $n=4$) is half-filled and determines the position of $E_{F}$. The
band gap of semiconducting states, which have spin in the direction opposite to that of
$m$-band, occurs between the filled flat $v_{1}$-band  and empty conduction $c_{1}$-band.
According to these bands, the equilibrium ballistic conductance of the infinite C$_{3}$Cr is
G$_\downarrow$=2$e^{2}/h$ for minority spin, but zero for majority spin. The calculated spin
projected total density of states (TDOS) in Fig. \ref{fig:band1} shows the energy spectrum of
majority and minority spin states in an interval $\pm$2 eV around $E_F$. The band gap for one
spin direction, and finite density of states at $E_{F}$ for the opposite spin are clearly
seen. This is a dramatically different finding than those of Ref.~\cite{dag,yang,rodrigues}.
Orbital projected local densities of states at Cr and C atoms show the orbital composition of
the spin-polarized bands. The $m$-band is composed of Cr-$3d$ and mainly first-neighbor
C-$2p$  orbitals at $E_{F}$, that is the $p-d$ hybridization. The flat $v_{1}$-band nearest
to $E_{F}$ is derived from the Cr-$3d$ and $4s$ states. The empty $c_{1}$-band originates
from C-$2p$ and Cr-$3d$ states.

The effect of spin-orbit (SO) coupling on the HM properties of C$_{3}$Cr has been calculated
by using all-electron DFT (Wien2K) code. We found the splitting is very small and
$E_{T}$(with SO)-$E_{T}$(without SO)=-7.9 meV. As illustrated in Fig. \ref{fig:band1}(c), the
difference between TDOSs calculated with and without SO-coupling is negligible and hence the
effect is not strong enough to destroy the half-metallic properties. This conclusion obtained
for $n=$3 can apply to other chain structures in Table I, since the Cr-C interaction decays
fast beyond the first nearest neighbors of Cr atom.

\begin{figure}
\includegraphics[scale=0.4]{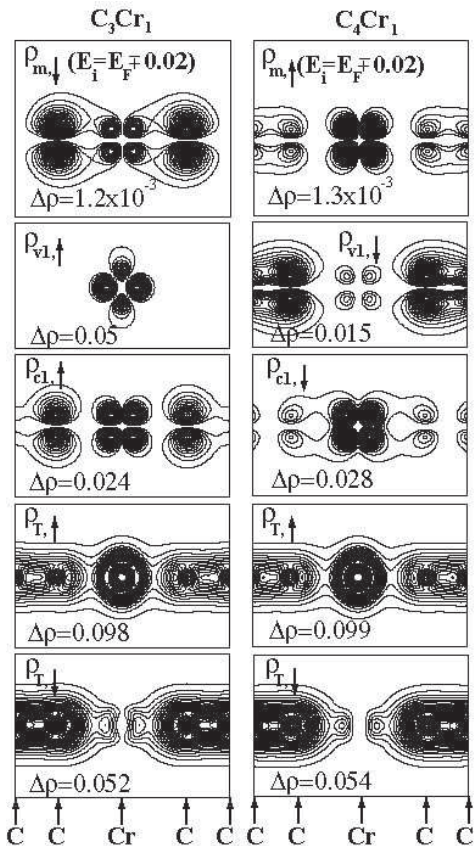}
\caption{ Charge density contour plots of linear chains of
C$_{3}$Cr, C$_{4}$Cr compounds on a plane through the chain axis.
$\rho_{m,\uparrow \rm{or} \downarrow}$ is the charge density of
metallic spin-states within energy range $E_{F}\pm0.02$ eV.
$\rho_{v_{1}, \uparrow {\rm or} \downarrow}$ and $\rho_{c_{1},
\uparrow {\rm or} \downarrow}$  are the charge density of the
highest valence $v_1$ and  the lowest (empty) conduction band
$c_1$, respectively. $\rho_{T,\uparrow}$ and $\rho_{T,\downarrow}$
are total charge density due to majority spin states  and minority
spin states, respectively. $\Delta \rho$ is contour spacing.}
\label{fig:charge}
\end{figure}

Further insight about the electronic structure and the character of bonding can be gained by
examining the charge distributions associated with selected bands shown in
Fig.~\ref{fig:charge}. The charge density of the metallic spin-state at $E_{F}$ is obtained
by averaging charges of states of $m$-band having energy $\pm$0.02 eV around $E_F$. They are
formed from the bonding combination (or $p-d$ hybridization) of C-$2p_{x,y}$ and
Cr-$3d_{xz,yz}$ valence orbitals of C$_{3}$Cr. In the case of even $n$ ($n$=4) it corresponds
to an antibonding combination of the above orbitals with enhanced Cr-$3d$ contribution.  The
charge density of $v_{1}$-band is  due to non-bonding Cr-$4s-3d_{z^{2}}$ orbitals in
C$_{3}$Cr. For even $n$ case, the C-$2p_{x,y}$ contribution is pronounced. Charge density of
the $c_{1}$-band suggests the antibonding combination of $p-d$ hybridized states.

The band structure and charge density plots suggest that the $p-d$
hybridization between neighboring C and Cr orbitals and resulting
exchange splitting of bands in different spin directions give rise
to the ferromagnetic ground state of the above chains. An
additional ingredient, namely cylindrical symmetry of the
$\pi$-bonds in the carbon chain provides conditions to result in
integer number of excess spin in the unit cell which is required
to achieve half-metallic behavior. The ferromagnetic ground state
with $\mu$ values integer multiple of $\mu_B$  per unit cell can
be understood from a local point of view based on first Hund's
rule. We take C$_3$Cr as examples. In Fig.~\ref{fig:band1} (a),
three spin-up (one non-degenerate-$v_1$, one doubly degenerate)
bands below $E_F$ are derived mainly from Cr-$3d$ orbitals. Five
of six electrons on Cr (in these corresponding bands) occupy the
majority spin states to yield N$_\uparrow=5$. The sixth electron
occupies the $p$-$d$ hybridized, doubly degenerate but only
half-filled $m\downarrow$-band yielding N$_\downarrow=1$.
Consequently, N=N$_\uparrow$$-$N$_\downarrow$=4, and hence
$\mu=4\mu_{B}$.

The present study predicts that linear chains of C$_{n}$(TM) compounds (TM=Cr,Ti,Mn,Fe,Co)
with specific $n$ can show half-metallic behavior with a diversity of spin-dependent
electronic properties.  Here, the type and number of atoms in the compound, as well as
even-odd $n$ disparity are critical variables available to engineer nanostructures with
spin-dependent properties. The electronic transport properties and the value of $\mu$ can be
modified also by applied axial strain.  Not only periodic structures, but also non-periodic
combinations comprising HM-HM or HM-S (or M) quantum structures and superlattices can be
envisaged to obtain desired device characteristics, such as spin-valve effect and
spin-resonant tunnelling.  Since linear carbon chains have been obtained also at the center
of multi-wall carbon nanotubes \cite{roth}, C$_{n}$Cr chains can, in principle, be produced
inside a nanotube to protect the spintronic device from the undesired external effects or
oxidation. In fact, we obtained that strained C$_{7}$Cr compound chain placed inside a (8,0)
SWNT can be a HM. Of course, the properties revealed in this study correspond to idealized
infinite chain structures, and  are subject to modifications when the chain size becomes
finite. However, for finite but long chains (for example a coil of C$_{n}$TM around an
insulating SWNT), the level spacings are still small to gain a band-like behavior. Also,
localization of electronic states due to imperfections in 1D may not lead to serious
difficulties when the localization length $\xi$ is larger than the length of the device. It
is also noted that the properties of chains may depend on the type and detailed atomic
structure of the electrodes.

In conclusion, we showed that half-metallic properties can be
realized in linear chains of carbon-transition metal compounds
presenting a number of exciting properties which can be of
fundamental and technological interest for new generation devices.
We believe that in view of recent progress made in synthesizing C
atomic chains, present study will bring a new perspective in
spintronics.


\begin{references}
\bibitem{spinrel}
Spin-relaxed structure optimization calculations starting with an
initial net spin $S_{T}=0$ have converged to the values
$S_{T}$=$\sum_{n,k}^{occ}$($S_{n,k,\uparrow}$+$S_{n,k,\downarrow}$)$>0$.
This situation  excludes the possibility of an antiferromagnetic
ground state.
\bibitem{energy} C$_{5}$+Cr initial state is a bound state with $\mu$=5.1 Bohr magneton and
high-spin polarization. We estimate the breaking strain of the linear C$_{5}$Cr HM at the
axial strain of $\epsilon \sim 0.1$ near $E_{T}(c)\sim 0$ in Fig.~\ref{fig:trans}(a).
\end{references}
\end{document}